\documentclass[11pt, final,3p,times]{elsarticle}
\pdfoutput=1
\usepackage[dvipsnames,svgnames,x11names]{xcolor}
\usepackage[final]{changes}

\usepackage{lineno,hyperref}
\usepackage{amsmath,amssymb,amsfonts}
\usepackage{algorithmic}
\usepackage{graphicx}
\usepackage{textcomp}
\usepackage{todonotes}
\usepackage[utf8]{inputenc}
\usepackage{mathtools, nccmath}
\usepackage{subcaption}
\usepackage{booktabs}
\usepackage{cleveref}
\usepackage{xcolor}
\usepackage{ucs}
\usepackage{algorithm,algorithmic}

\usepackage[labelfont=bf]{caption}
\modulolinenumbers[1]

\journal{Transportation Research Part C: Emerging Technologies}

\biboptions{authoryear}

\begin{document}

\begin{frontmatter}

\title{Deep convolutional generative adversarial networks for traffic data imputation encoding time series as images}


\author[mymainaddress]{Tongge Huang} 

\author[mysecondaryaddress]{Pranamesh Chakraborty}

\author[mymainaddress]{Anuj Sharma}

\address[mymainaddress]{Civil, Construction, and Environmental Engineering Department, Iowa State University, Ames, Iowa, USA 50011}
\address[mysecondaryaddress]{Department of Civil Engineering, Indian Institute of Technology Kanpur, Kanpur, U.P, INDIA 208016}

\begin{abstract}

Sufficient high-quality traffic data are a crucial component of various Intelligent Transportation System (ITS) applications and research related to congestion prediction, speed prediction, incident detection, and other traffic operation tasks. Nonetheless, missing traffic data are a common issue in sensor data which is inevitable due to several reasons, such as malfunctioning, poor maintenance or calibration, and intermittent communications. Such missing data issues often make data analysis and decision-making complicated and challenging. In this study, we have developed a generative adversarial network (GAN) based traffic sensor data imputation framework (TSDIGAN) to efficiently reconstruct the missing data by generating realistic synthetic data. In recent years, GANs have shown impressive success in image data generation. However, generating traffic data by taking advantage of GAN based modeling is a challenging task, since traffic data have strong time dependency. To address this problem, we propose a novel time-dependent encoding method called the Gramian Angular Summation Field (GASF) that converts the problem of traffic time-series data generation into that of image generation. We have evaluated and tested our proposed model using the benchmark dataset provided by Caltrans Performance Management Systems (PeMS). This study shows that the proposed model can significantly improve the traffic data imputation accuracy in terms of Mean Absolute Error (MAE) and Root Mean Squared Error (RMSE) compared to state-of-the-art models on the benchmark dataset. Further, the model achieves reasonably high accuracy in imputation tasks even under a very high missing data rate ($>$ 50\%), which shows the robustness and efficiency of the proposed model. 

\end{abstract}


\end{frontmatter}

\section{Introduction}\label{sec:intro}

Dissemination of accurate traffic data is an essential requirement for supporting advanced traffic management system operations. Different types of sensors, such as loop detectors, radar sensors, and video detectors, are installed in freeways and arterials for traffic data collection purposes. The data collected from these sensors can be used to detect traffic congestion or incidents \citep{chakraborty2018traffic, chakraborty2018freeway, chakraborty2019data}, provide travel time information to road users \citep{lu2017real, gan2017estimation}, and support decision making at the traffic operation and planning levels \citep{shi2015big, ma2017traffic}. However, missing data are quite common in traffic sensor data due to issues such as malfunctioning, poor maintenance or calibration, and intermittent communications \citep{lee2011quantifying}. According to the California Performance Measurement System (PeMS), only 67\% of the sensors in District 7 of southern California (Los Angeles) were found to be working as expected in December 2018 \citep{caltrans2019}. While data from sensors with a high percentage of missing data can be discarded from further usage, an alternate approach is to impute the missing records, so that these sensors’ data can still be used for subsequent analysis. This is particularly important for traffic-related studies that require traffic records to be complete, such as traffic flow analysis methods. Therefore, it is important to develop an effective traffic data imputation method which can handle missing traffic records even at a high percentage of missing data.

Traditionally, traffic data imputation has been done using prediction or interpolation methods that use historical traffic data or traffic data from adjacent sensors or time points to impute missing records \citep{nihan1997aid, ghosh2007bayesian, allison2001missing, chang2012missing}. However, these methods often fail to explicitly capture the spatio-temporal variations, which can lead to unreliable performance. Another class of imputation techniques relies on statistical learning models such as Markov chains or principal component analysis to learn the schema of the traffic data matrix \citep{lv2014traffic, ni2005markov, qu2009ppca}. However, these methods require the assumptions on the probability distribution of traffic data, which makes them difficult to apply in real-world scenarios. Also, these methods do not work well when handling large proportions of missing data, which is a common issue in real-world too. With the recent advancements in deep learning techniques and their success in image recognition and imputation tasks, traffic data imputation problem has been tackled using these new techniques that treat the data imputation problem as a corrupted data denoising problem \citep{duan2016efficient, ku2016clustering, asadi2019convolution}. However, modeling the strong time dependency of the time-series data is one of the major challenges in the application of these imputation techniques. To address this issue, we propose a novel Gramian Angular Summation Field (GASF) encoding method in this study to embed the traffic data for our model input, precisely preserving its time dependency. We then train a deep convolutional generative adversarial network (DCGAN) to generate realistic synthetic data for missing data imputation.

In recent years, deep learning based Generative Adversarial Networks (GANs) have been successful in generating impressively realistic synthetic data by modeling the real data distributions \citep{goodfellow2014generative, goodfellow2016deep}. Further, by taking advantage of convolutional neural networks (CNNs), DCGANs \citep{radford2015unsupervised} have shown remarkable ability in generating high quality synthetic image data for many applications such as image-to-image translation \citep{isola2017image}, audio generation \citep{donahue2018adversarial}, and image super-resolution \citep{ledig2017photo}. Such impressive performance in modeling the original data distribution has made DCGANs a strong candidate for data imputations \citep{yeh2017semantic, lee2019collagan}.

In this study, we have developed a traffic data imputation framework based on generative adversarial network (TSDIGAN) to efficiently resolve the missing data problem. Our proposed model treats the data imputation problem as a synthetic data generation problem. The novel GASF encoding method used in this study helps to embed the strong temporal dependency of the time-series data, thereby translating the time-series imputation problem to an image imputation problem. We evaluate our proposed model using the benchmark PeMS dataset \citep{PeMS} and compare its performance with other baseline statistical and deep learning models. We also investigate the capability of our proposed model for large-scale applications by clustering sensors into homogeneous groups and learning imputation models for each cluster of sensors. Thus, the major contributions of our study are as follows:

\begin{itemize}

    \item Our proposed model takes advantage of deep learning based generative models, enabling users to treat the data imputation problem as a data generation problem. Such a generative framework can impute the missing data using the best-fitting generated realistic looking data, such that it is adaptive and robust in imputing the missing records, even at a high percentage of missing data.
    
    \item Our novel traffic time-series data-encoding technique using GASF method preserves the time dependency of traffic data without losing the underlying temporal dependency information. This proposed encoding method helps the model to learn the point-wise temporal relations between time-series traffic data.
    
    \item Our proposed model achieved reasonably high accuracy in imputation task for missing data ratios ranging from 5\% to 90\%, making it robust and reliable under challenging high missing data percentages. Additionally, training the proposed model using year-long traffic data takes less than 8 minutes, making it efficient and scalable for large-scale implementation. Therefore, we also evaluated our proposed model across extensive sensor groups of California District 5, showing the feasibility for large-scale practical applications. The proposed model has been found to improve the imputation accuracies in terms of Mean Absolute Error (MAE) and Root Mean Squared Error (RMSE) compared to the state-of-the-art benchmark imputation models, while achieving comparable results in terms of Mean Relative Error (MRE).

\end{itemize}

The rest of this paper is organized as follows. Section \ref{sec:lit_review} provides a brief description of related work regarding traffic missing data imputations, followed by the details of our methodology in Section \ref{sec:methodology}. Section \ref{sec:results} provides a detailed description of the data used in this study, results obtained using our proposed model compared to the baseline models. Finally, Section \ref{sec:conclusion} summarizes the contributions of our study and its implications for future research.

\section{Literature Review}\label{sec:lit_review}

Since several ITS applications require high-quality traffic data, a significant amount of studies have been done in the past on missing traffic data imputation. As summarized by \cite{li2014missing}, traffic data imputation methods can be broadly divided into three categories: prediction methods, interpolation methods, and statistical learning methods. Some recent studies have also applied deep learning-based methods like stacked denoising autoencoders (DSAE) to estimate missing traffic data.

Prediction based models such as the Autoregressive Integrated Moving Average (ARIMA) model \citep{nihan1997aid, park1998short}, support vector regression (SVR) \citep{castro2009online}, and Bayesian networks (BNs) \citep{ghosh2007bayesian} use the historical observations to predict the future data points. These prediction-based models assume that the future data points follow the typical trace as the historical data. Prediction based methods can usually estimate data points effectively over the short-term and for samples whose missing data ratio is low. However, their performance drops significantly for long-term imputation problems. Further, these methods only use observations from the history before the missing data points, while any valuable information available after the missing data points is not utilized.

Interpolation based models, another popular method for missing data imputation, include temporal neighboring model or historical model \citep{yin2012imputing, smith2003exploring, allison2001missing} and k-NN model \citep{al2004new}. The basic idea of the historical model is to impute missing data points using the average historical value reported by the same sensor for the same time periods. Therefore, historical models assume that the daily traffic flow has strong consistency. However, they do not use the information of the inherent daily variation to improve the imputation performance. In contrast, the k-NN method utilizes the average historical value for a given day of the week from the same sensor or neighboring sensors to impute missing data, instead of considering the overall average. However, these interpolation-based models often fail to explicitly capture the spatial-temporal variations, which can lead to unreliable imputation results.

Statistical based models such as probabilistic principal component analysis (PPCA) \citep{qu2009ppca} and Markov Chain Monte Carlo (MCMC) methods \citep{ni2005markov} have been proposed to overcome the limitations mentioned above and improve imputation accuracy based on statistical modeling. These statistical methods assume a probability distribution model for the traffic data and impute the missing data using the observed data with optimized parameters. However, these models do not work well when the proportion of missing data are significantly high, especially for the case when an entire day of traffic data is missing \citep{anandkumar2014tensor, tan2013tensor}, a situation fairly common in real-world scenarios.

In recent years, deep learning based models using denoising stacked autoencoders (DSAE) have been studied to overcome the shortcomings of the traditional data imputation models \citep{duan2016efficient, ku2016clustering}. Such models have been found to be successful in obtaining reliable performance by converting the data imputation problem into a data cleaning/denoising problem. Typically, DSAE models extract the useful inherent correlations from the original data, recovering them from the high-level features with noise reduction. The basic idea of DSAE models is to train a ``recovery tool" using both the raw data and imputed missing data. Therefore, a well-trained model will recover missing data with more reliable estimations compared to the conventional methods described above. However, such feature extraction compresses data into lower dimensions, which limits the variability of the model and makes model outcomes less interpretable. 

\textcolor{black}{Besides this, recurrent neural network (RNN) based GAN has also been used for time series data imputation. For instance, conditional Long Short-Term Memory (LSTM) based GAN has been used for medical data generation \citep{esteban2017real} and traffic data prediction \citep{lv2018generative}. On the other hand, Gated Recurrent Unit (GRU) based GAN has been used for multivariate time series generation \citep{luo2018multivariate}. Similarly, \cite{asadi2019convolution} adopted convolution recurrent autoencoder using bidirectional-LSTM layer as the encoder layer for spatial-temporal missing data imputation.  However, training such RNN networks generally take significantly longer time, particularly when handling the long time sequence ($>$200) \citep{li2019mad}.  Recently, \cite{chen2019traffic} proposed parallel data based GAN model, which used the real data and synthetically generated data simultaneously for traffic data imputation to achieve state-of-the-art results. However, using the original daily traffic time series data as the latent space limits the generative ability of GAN based model thereby requiring each sensor to have its own generative model. This leads to training and managing a large number of models thereby making it difficult for large-scale  application.}

\textcolor{black}{In this study, we propose a traffic data imputation framework based on generative adversarial network (TSDIGAN)  encoding the time series into images. This enables us to treat the data imputation problem as image generation problem, thereby utilizing the significant developments in DCGAN based image generation problems. We also compare our proposed model performance with the most recent state-of-the-art traffic data imputation based on GAN, proposed by \cite{chen2019traffic} to show the efficacy of our proposed model. }

\section{Methodology}\label{sec:methodology}

In this section, we provide the step by step details of our proposed TSDIGAN framework. We first explain the notation used in this study and then introduce the Gramian Angular Summation Field (GASF) time-series encoding and the basic concepts of GAN. The details of the proposed TSDIGAN model framework is then discussed, followed by the large-scale implementation technique.

\subsection{Notation}\label{sec:notation}

We first describe the abstract mathematical expressions used in this study. These notations will be used throughout the paper by default. Let us assume that the traffic flow time-series data for a given sensor $s$ on a given day $d$ is denoted by $X^s_d = \{{x_1},{x_2},{x_3},...,{x_t},...,{x_T}\}$, where $x_t$ is the $t^{th}$ observation of $X$. For 5-minute interval traffic flow data used in this study, $T$ is given by $T = \{{t_i}\}^{288}_{i=1}$. The combined daily time-series data for each sensor $s$ can then be represented as ${\tilde X^s}=\{X^s_{d}\}_{d=1}^{D}$, where $D$ represents the total number of days involved. Finally, we use a binary corrupted mask as an indicator variable to flag whether the data for $X^s_{d,t}$ is present or missing. This leads to Equation \ref{eq:indicator}.

\begin{equation}\label{eq:indicator}
{I^s_{d,t}} = \left\{ {\begin{array}{*{20}{ll}}
{{0}, \text{    {if }}{{{X}}^s_{{{d,t}}}}{\text{ is missing}}} \\ 
{{1}, \text{    {if }}{{{X}}^s_{{{d,t}}}}{\text{ is not missing}}}
\end{array}} \right\}
\end{equation}

${\tilde X^s}$ can be divided into two subsets: (1) fully observed datasets, which do not have missing data points in any samples, denoted as ${\tilde X^{s,f}}=\{X^{s,f}_d\}_{d=1}^{D_1}$, and (2) corrupted datasets, which contain missing data points in each samples, denoted as ${\tilde X^{s,m}}=\{X^{s,m}_d\}_{d=1}^{D_2}$. For each sample, we flag whether the data points are missing or not using the corrupted mask $I^s_{d,t}$ which can also be divided into two subset matrices: $I^{s,f}_{d,t}$ (for fully observed datasets) and $I^{s,m}_{d,t}$ (for corrupted datasets). Next, we describe the first task in our proposed TSDIGAN framework: converting time-series traffic data to GASF encoding, which enables treatment of the time-series imputation problem as an image imputation problem.

\subsection{Gramian Angular Summation Field}\label{sec:gasf}

\textcolor{black}{Encoding time-series data as images have been widely used for time-series classification, audio data recognition, and similar other tasks. One of the popular approaches to tackle this problem is the spectrogram based method. For example, \cite{cummins2017image, zhao2018sound} converted the speech/sound time series data into spectrogram using Short-time Fourier Transform (STFT) for recognition of emotional speech and locate image regions which produce sounds. Also, \cite{lefebvre2017traffic} estimated traffic flow by converting the spectrum features of the acoustic sensors signal data using Mel Frequency Cepstral Coefficients (MFCCs). However, such spectrogram based methods require careful parameter selection for precise inverse operation, and the imputation task of typical daily traffic volume data is unlikely to benefit from it  \citep{wang2015encoding}. Another approach to encode the time series to images is to combine the spatial-temporal information as a 2D matrix. For instance, \cite{zhuang2018innovative, kim2018capsule} merged the ordered road segments/stations and time-series traffic data to form a 2D matrix that can be used by CNN to extract the spatial-temporal information. However, obtaining well-ordered sensor information in a complex, large-scale network is difficult and time-consuming. Further, the model needs to be retrained completely even when a single sensor is added or removed to the network.}

 \textcolor{black}{To alleviate these issues, in this study, we adopted the Gramian Angular Summation Field (GASF), which has been demonstrated to improve CNN features extraction \citep{wang2015encoding, wang2017time}. The GASF has the following advantages. First, it helps to preserve and enhance the temporal correlations by considering the trigonometric sum between each time instance point. Second, the character of bijection provides directly and precisely inverse operation without the need for any specific parameter selection.} 

We rescaled the given preprocessed traffic flow time-series data ${\tilde X^{s,f}}$ to $[0, 1]$ such that we can represent the data in polar coordination system. Therefore, for daily traffic time-series data $X^{s,f}_d = \{{x_1},{x_2},...,{x_t},...,{x_T}\}$ , the volume data $x_t$ and  time stamp $t_i$ are encoded as angular cosine and radius ($r$) respectively,  given by the following equation:

\begin{equation}\label{eq:gasf_polar}
\begin{cases}\phi = \arccos ({{x}_t}),0 \le {{x}_t} \le 1\\
r = \frac{{t_i}}{C},{t_i} \in {\mathbb N}
\end{cases}
\end{equation}
where, $C$ is a constant regularization factor.

After transforming the traffic time-series data by using the above equation, the temporal correlation between each point can be identified by the GASF matrix denoted as ${\hat X^{s,f}}_{d}$:

\begin{equation}\label{eq:gasf_transform}
{\hat X^{s,f}}_{d} = \left[ {\begin{array}{*{20}{c}}
  {\cos ({\phi _1} + {\phi _1})}& \cdots &{\cos ({\phi _1} + {\phi _T})} \\
   \vdots & \ddots & \vdots  \\ 
  {\cos ({\phi _T} + {\phi _1})}& \cdots &{\cos ({\phi _T} + {\phi _T})} 
\end{array}} \right]
\end{equation}

The main diagonal of ${\hat X^{s,f}}_{d}$ contains the original angular/value information. As mentioned above, the rescaled time-series data $X^{s,f}_d$ belongs to the set $[0,1]$ so that the mapping between $x_t$ and its corresponding angular cosine value is bijective. These characteristics allow us to precisely recover (inverse transform) the traffic time-series data from the GASF matrix ${\hat X^{s,f}}_{d}$ using the following equation:

\begin{equation}\label{eq:gasf_inverse_transform}
{X^{s,f}_d} = \sqrt {\frac{{{\hat X^{s,f}}_{d,diagonal} + 1}}{2}} = \sqrt {\frac{{\cos (2\phi ) + 1}}{2}}\\, \phi  \in [0,\frac{\pi }{2}]
\end{equation}

Here, we replace the suspect 0 value with 1 to avoid the zero division error and apply  log transformation to avoid skewed distribution. Then, we rescale and transform the preprocessed 1-D daily traffic flow data $X^{s,f}_d$ into image-like GASF matrix ${\hat X^{s,f}}_{d}$ using Equations \ref{eq:gasf_polar} and \ref{eq:gasf_transform}. The image-like matrix embedding with temporal dependencies helps convolutional neural networks (CNNs) to effectively extract the required features \citep{lecun1998gradient}. We used this image-like GASF matrix ${\hat X^{s,f}_d}$ as the input of our proposed GAN model. Finally, the daily traffic flow data $X^{s,f}_d$ can be recovered from the GASF matrix ${\hat X^{s,f}_d}$ using Equation \ref{eq:gasf_inverse_transform}.

\subsection{Generative Adversarial Networks}\label{sec:gan_description}
\begin{figure}[!htb]
    \centering
    \includegraphics[width=0.75\textwidth]{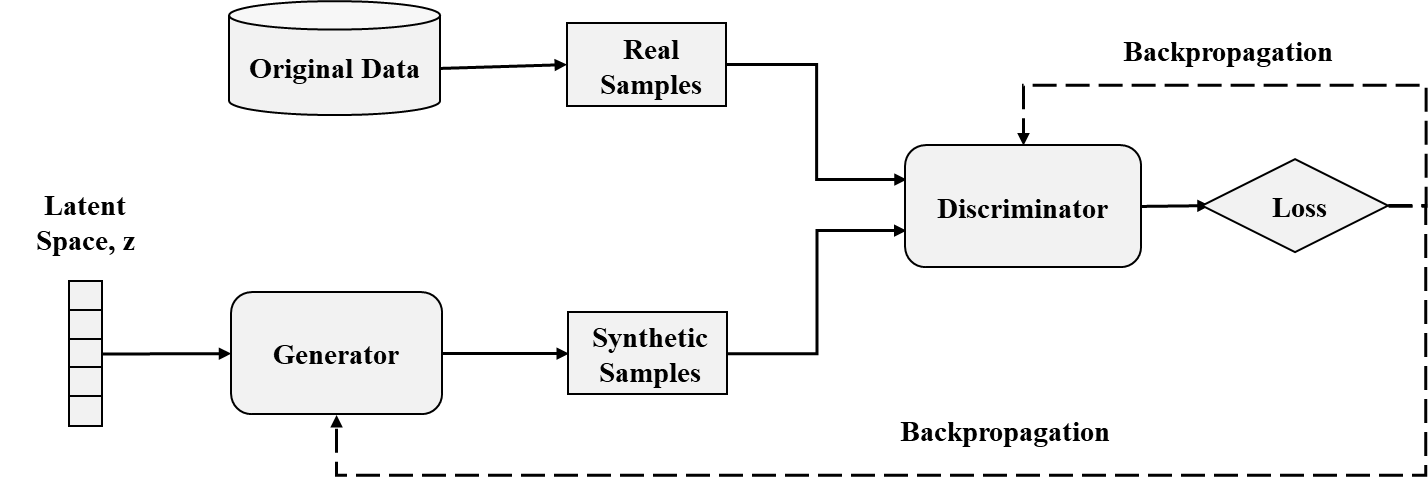}
    \caption{Architecture of GANs.}
    \label{fig:gan_structure}
\end{figure}

Generative adversarial nets (GAN) were introduced as an effective tool for data augmentation and data generation. The basic GAN architecture is shown in Figure \ref{fig:gan_structure}, which consists of two parts: a generator and a discriminator \citep{goodfellow2014generative}. The generator $(G)$ takes a random vector $z$ as input, sampled from a noise distribution $p_z$,  to output a corresponding synthetic data sample $G(z)$. The discriminator $(D)$ takes a real sample $x$ from the original dataset $p_{data}$ and the synthetic sample $G(z)$ as inputs to estimate the probability that the generated and real sample comes from the same distribution. In our case, the original dataset $p_{data}$ and the real sample $x$ is given by  ${\tilde X^{s,f}_d}$ and $\hat X^{s,f}_d$ respectively, which are obtained from GASF encoding. Both $G$ and $D$ usually consist of multi-layer perceptions (MLPs).

The discriminator and generator compete with each other like a two-player minimax game, where both the discriminator and the generator are trained simultaneously.  During the training, the generator tries to fool the discriminator, while the discriminator tries to distinguish the real samples $x$ from the synthetic samples $G(z)$ by solving the value function $V(G,D)$ formulated as:

\begin{equation}\label{eq:GAN_D}
\mathop{\min}\limits_G\mathop{\max}\limits_D V(D, G) = {E_{x{\sim}{p_{data}}(x)}}[\log D(x)] + {E_{z{\sim}{p_z}(z)}}[\log (1 - D(G(z)))]
\end{equation}

After alternative training of both the discriminator and generator, the distribution of synthetic samples $p_{syn}$ produced by the generator converges with the distribution of the original real data $p_{data}$. In other words, the generator produces such realistic synthetic samples that the discriminator can no longer distinguish them from the original data samples.

\subsection{TSDIGAN Architecture}\label{sec:TSDIGAN_structure}
\begin{figure}[htbp]
    \centering
    \includegraphics[width=0.85\textwidth]{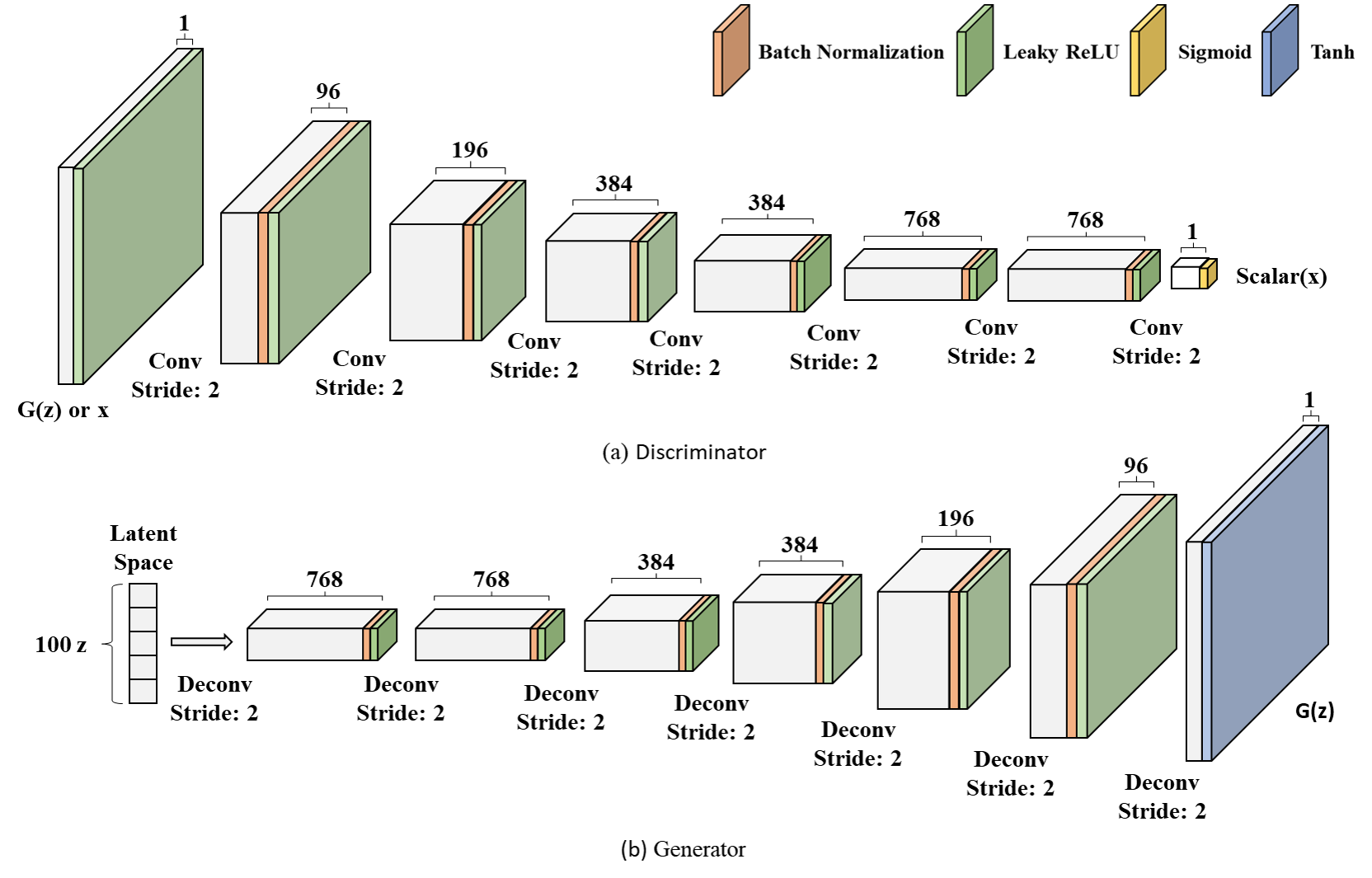}
    \caption{Proposed TSDIGAN (a) Discriminator and (b) Generator.}
    \label{fig:TSDIGAN_cnn}
\end{figure}

In this subsection, we introduce the framework of our proposed model in details. The architecture of our proposed discriminator and generator is shown in Figure \ref{fig:TSDIGAN_cnn}. 

In the discriminator module, we used strided convolution (expressed as ``Conv Stride: 2” in Figure \ref{fig:TSDIGAN_cnn}a). In contrast, for the generator module, we used fractional-strided convolutions (expressed as ``Deconv Stride: 2” in Figure \ref{fig:TSDIGAN_cnn}b). We then applied Batch Normalization (BatchNorm) layers to stabilize the learning process and prevent model collapse. BatchNorm was used for all layers, except the input layer of the discriminator and the output layer of the generator to avoid sample oscillation. We used LeakyReLU as the activation function for all layers to provide non-linearity, except in the output layers of both the discriminator and generator, where we used the sigmoid and tanh functions as the activation functions to produce the scalar and synthetic data respectively. Moreover, the latent random vector $z$ was sampled from the normal distribution. 

To help the CNN model learn more effectively, we repeated the first and last traffic flow data points at the head and tail of the $X^{s,f}_d$ three times  instead of doing the ``zero-padding". Therefore, the dimensions of the input traffic data vector transformed to 294 = $(3+288+3)$, with the GASF matrix $\hat X^{s,f}_d$dimensions being $294 \times 294$.  This same padding value was removed after training. Additionally, we applied a Gaussian filter on the $\hat X^{s,f}_d$ to reduce the inherent noise and improve the quality of the GAN-generated synthetic data \citep{susmelj2017abc}. The initial learning rate was set to 0.0002  along with the Adam optimizer. \textcolor{black}{
GANs are however known to suffer from mode collapse issues frequently, when the training model often sticks to only few modes of the true distribution ignoring the other modes \citep{radford2015unsupervised}. To prevent the potential mode collapse in this study, we randomly assigned the training label from 0.8 to 1.1 and 0.0 to 0.3 for positive and negative labels respectively. Also, we randomly flipped 10\% of the training labels in each mini-batch \citep{salimans2016improved}. These strategies helped to prevent the discriminator/generator from trapping into state of extremely high confidence and stabilize the training process during our experiments.} 

\subsubsection{Maximum Mean Discrepancy}\label{sec:mmd}
After training our TSDIGAN model, the generator can generate synthetic daily traffic flow dataset ${\tilde X^{s,syn}}=\{X^{s, syn}_d\}_{d=1}^{D_3}$ that looks ``close" to the real data ${X^{s,f}_d}$ sampled from the original fully observed dataset ${\tilde X}^{s,f}$. During the training phase, the generator generates the same number of samples as used for training. Therefore, in this case $D1$ is obviously equal to $D3$. In general, a well-trained GAN can implicitly learn the distribution of the original fully observed dataset ${\tilde X}^{s,f}$. Visual inspection of synthetic traffic flow data is one recommended means of determining if a TSDIGAN model is well-trained, which is also considered as the intuitive way to inspect GANs based models \citep{borji2019pros}. 

In addition, we utilized the maximum mean discrepancy (MMD) method as the quantifiable tool to measure the similarity between the two distributions ${\tilde X^{s,f}}$ and ${\tilde X^{s,syn}}$ \citep{gretton2007kernel} using the following equation: 

\begin{equation}\label{eq:mmd}
\widehat{MMD_{u}} = {\left[ {\frac{1}{{{D_1}^2}}\sum\limits_{i,j = 1}^{{D_1}} {k(\tilde X_i^{s,f},\tilde X_j^{s,f})}  - \frac{2}{{{D_1}{D_3}}}\sum\limits_{i,j = 1}^{{D_1},{D_3}} {k(\tilde X_i^{s,f},\tilde X_j^{s,syn})}  + \frac{1}{{{D_3}^2}}\sum\limits_{i,j = 1}^{{D_3}} {k(\tilde X_i^{s,syn},\tilde X_j^{s,syn})} } \right]^{\frac{1}{2}}}\
\end{equation}

\begin{figure}[htbp]
    \centering
    \includegraphics[width=0.65\textwidth, height=0.18\textwidth]{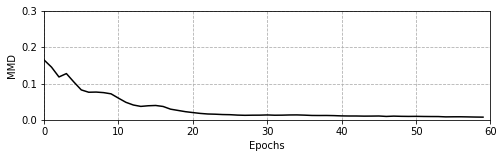}
    \caption{MMD Score Trace}
    \label{fig:mmd}
\end{figure}

Here, $k(\tilde X_i^f,\tilde X_j^{syn})$ represent the kernel function, and we used the radial basis function (RBF) kernel for MMD score calculation as described in \cite{esteban2017real}. We recommend interested readers refer to \cite{esteban2017real} for more details. Figure \ref{fig:mmd} shows the sample MMD score trace over 60 training epochs. Therefore,  training of the proposed TSDIGAN  was verified not  only through visual inspection of its synthetic traffic flow data, but also by observing stable convergence of the $MMD$ score.

\subsection{TSDIGAN Imputation Model}\label{sec:TSDIGAN_imputation}

\begin{figure}[htbp]
    \centering
    \includegraphics[width=0.85\textwidth]{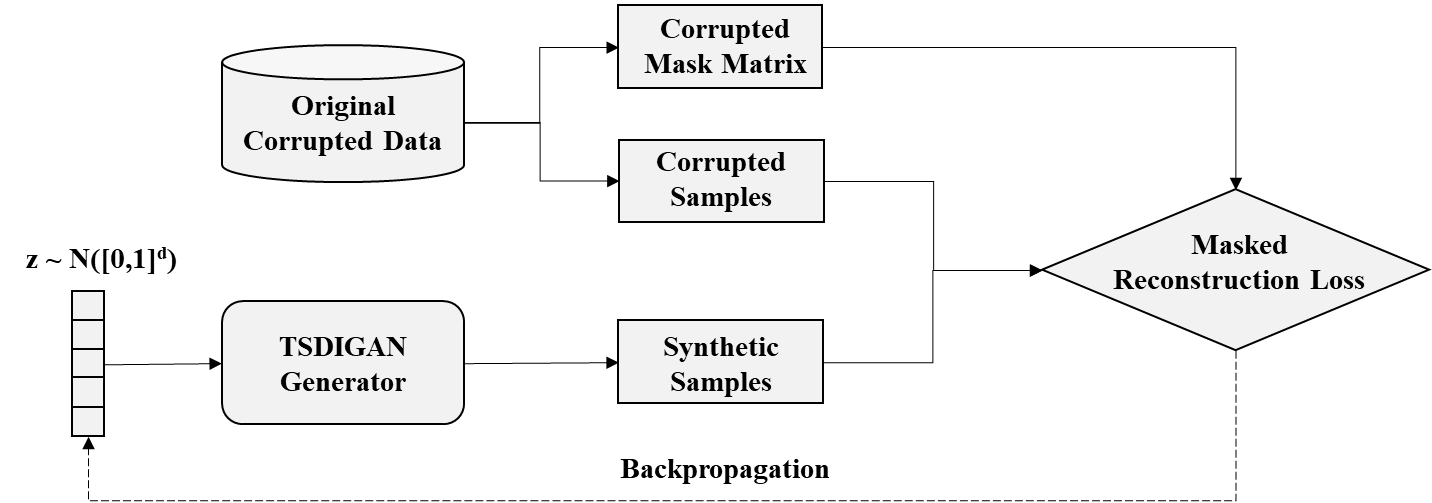}
    \caption{Architecture of TSDIGAN Imputation Model.}
    \label{fig:TSDIGAN_imputation}
\end{figure}

After training our TSDIGAN using the fully observed dataset ${\tilde X^{s,f}}$ as described in Section \ref{sec:TSDIGAN_structure}, we used our trained model for missing traffic data imputation. In this subsection, we introduce the imputation (or inpainting) framework shown in Figure \ref{fig:TSDIGAN_imputation}. The basic idea of our imputation framework is similar to GAN based image inpainting \citep{yeh2017semantic, yu2018generative} in that we searched the most representative $z$ from $p_z$ as the input for the generator to use in generating a realistic synthetic data for each specific ${X}^{s,m}_d$.

From the basic concept of the GAN, we know that the generator learns the distribution from original fully observed dataset ${\tilde X}^{s,f}$, and generate the synthetic dataset ${\tilde X}^{s,syn}$ without any missing points from the learned distributions. We could thus use the synthetic data sample ${X}^{s,syn}_d$ to fill in the missing points in ${X}^{s,m}_d$ via the corrupted mask vector $I^{s,m}_{d,t}$. However, for the given incomplete data vector ${X}^{s,m}_d$, which $z$ should we use to generate the most reasonable data? The question above can be described as searching for the ``closest" $z_c$ from $p_z$ to generate missing values constituting the best ``overlay" synthetic traffic flow data ${X}^{s,syn}_d$, based on the observed part of the given data vector ${X}^{s,m}_d$. To accomplish this goal and inspired by \cite{yeh2017semantic, luo2018multivariate}, we designed the masked reconstruction loss denoted as ${\ell_r}$. For any given ${X}^{s,m}_d$, the masked reconstruction loss can be formulated as: 

\begin{equation}\label{eq:maskreconstructionloss}
{\ell _r}({z}{|}{X^{s,m}_d}) = \frac{{\left\| {X_d^{s,m} \odot I_d^{s,m} - X_d^{s,syn} \odot I_{d}^{s,m}} \right\|}_1}{{\sum\limits_{t = 1}^T {I_{d,t}^{s,m}}}}
\end{equation}

It should be noted that by multiplying the masked missing vector $I^{s,m}_d$, only the observed part of ${X}^{s,m}_d$ is used for calculating the masked reconstruction loss, and the $\odot$ represents element-wise multiplication. Thus, the ``closest" latent vector $z_c$ can be represented as:

\begin{equation}\label{eq:bestz}
{z_c} = \mathop {\arg \min }\limits_z {\ell _r}({z}{|}{X^{s,m}_{d}})
\end{equation}

Finally, we used the synthetic traffic flow data ${X}^{s,syn}_d$ generated from the ``closest" $z_c$ to fill in the missing part of ${X}^{s,m}_d$. We summarize this imputation module step by step as shown in Algorithm \ref{alg:imputation}.

\begin{frame}

\begin{algorithm}[H]
\begin{algorithmic}[1]

\REQUIRE{(1) Corrupted traffic data vector $X^{s,m}_d = \{{x_1},{x_2},{x_3},...,{x_t},...,{x_T}\}$ which need imputation, (2) Corrupted missing vector $I^{s,m}_{d}$, (3) number of iterations $w$ for back-propagation on latent space $z$, (4) learning rate $\alpha$ and (5) trained generator.}
\newline
\STATE Initialize $z \sim N([0,1])$ as the input for generator.
\STATE Fix the weights of trained generator and active the gradient descent on $z$ 
\FOR{$w \space $ iterations}

\STATE Generate GASF matrix from the $z$, and extract the synthetic traffic time-series data using Equation \ref{eq:gasf_inverse_transform}:

\begin{align*}
\hat X^{s,syn}_d = G(z)
\end{align*}

\begin{align*}
X_d^{s,syn} = \sqrt {\frac{{diag(\hat X_d^{s,syn}) + 1}}{2}}
\end{align*}

\STATE Calculate masked reconstruction loss, and apply Back-propagation on $z$: 

\begin{align*}
{\ell _r}({z}{|}{X^{s,m}_d}) = \frac{{\left\| {X_d^{s,m} \odot I_d^{s,m} - X_d^{s,syn} \odot I_{d}^{s,m}} \right\|}_1}{{\sum\limits_{t = 1}^T {I_{d,t}^{s,m}}}}
\end{align*}

\begin{align*}
z \leftarrow z - \alpha \times \nabla {\ell _r}({z}{|}{X^m_d})
\end{align*}

\STATE Obtain the ``closest" $z_c$ by:
\begin{align*}
{z_c} = \mathop {\arg \min }\limits_z {\ell _r}({z}{|}{X^{s,m}_{d}})
\end{align*}

\ENDFOR
\STATE Impute the ${X}^{s,m}_d$ using the $X^{s,syn}_d$ generated from $z_c$:
\begin{align*}
X_d^{s,\text {imputed}} = X_d^{s,syn} \odot (1 - I_d^{s,m}) + X_d^{s,m} \odot I_d^{s,m}
\end{align*}

\end{algorithmic}

\caption{Traffic data imputation module using TSDIGAN}
\label{alg:imputation}
\end{algorithm}
\end{frame}

\subsection{Large-Scale  Implementation}\label{sec:TSDIGAN_state}
Deep learning models are promising for the traffic data imputation task; however, their practical applications on large-scale statewide level requires further investigation. \textcolor{black}{In this subsection, we investigate the capability of our proposed TSDIGAN model for large-scale real world application. Typically,   traffic sensors over extensive wide coverage involves a wide variation of traffic data characteristics. Grouping all the sensors together in a single cluster can make the model training significantly harder due to multiple modes present in the data generated from the distinct variations in traffic data generated across the different sensors. Further, this can lead to model instability and mode collapse, thereby making the training process significantly difficult and lead to poor model performance. On the other hand, training a model individually for each sensor leads to a significantly larger number of models training and maintenance/updates, making it difficult to large-scale application. For example, in the recent state-of-the-art traffic data imputation study by Chen et al. (2019) using parallel data based GAN model, a total of 294 models across the 147 districtwide sensors were developed for weekday and non weekdays. In this study, we chose a middle ground where we group the sensors based on their inherent traffic data characteristics such that models generated for each cluster can be focused towards the cluster-specific traffic variation characteristics. More specifically, we use k-means clustering \citep{macqueen1967some, berkhin2006survey} to group the sensors based on their daily traffic flow patterns because of it's simplified approach and computation efficiency.} 

Let us assume, we have $S$ sensors in the traffic sensor networks. As mentioned in Section \ref{sec:notation}, the fully observed data provided by a given sensor $s$ is denoted as ${\tilde X^{s,f}}=\{X^{s,f}_d\}_{d=1}^{D_1}$. Therefore, the traffic flow values for each sensor over a given set of days ($D_1$) can be denoted as a matrix with the dimensions $D_1 \times T$. At each time instance $t$, we extracted the features of this $D_1 \times T$ matrix by taking the quantiles $q$ value along the $D$, and augmenting them into one long feature vector $X^{s,f}_{feature}$ with the shape $1 \times 1440$ as:

\begin{equation}\label{eq:features}
{X^{s,f}_{feature}} = \left[ {q{\text{(}}{{\tilde X}^{s,f}}{\text{,}}10{\text{), }}q{\text{(}}{{\tilde X}^{s,f}}{\text{,}}30{\text{), }}q{\text{(}}{{\tilde X}^{s,f}}{\text{,}}50{\text{), }}q{\text{(}}{{\tilde X}^{s,f}}{\text{,}}70{\text{), }}q{\text{(}}{{\tilde X}^{s,f}}{\text{,}}90{\text{)}}} \right]
\end{equation}

We represent long feature vectors for all the sensors with $\tilde X^{s,f}_{feature} =\{X^{s,f}_{feature}\}_{s=1}^{S}$, and used the k-mean and elbow method \citep{kodinariya2013review} to divide the sensors into different groups. This sensor clustering procedure is summarized in Algorithm \ref{alg:kmean}. We then trained our proposed TSDIGAN model for each group separately. This enabled us to simply identify which group any sensor belonged to, and use its corresponding trained model to produce appropriate synthetic data for imputation.

\begin{frame}

\begin{algorithm}[H]

\textbf{Input}: Feature vectors for all the sensors $\tilde X^{s,f}_{feature} =\{X^{1,f}_{feature},X^{2,f}_{feature},...,X^{s,f}_{feature},...,X^{S,f}_{feature}\}$ \\
\textbf{Output}: Sensor groups
\begin{algorithmic}[1]

\FOR{i=1 to $S$ }
\STATE Initialize: $K$=i, $K$ clustering centroids $\mu_1, \mu_2 \in \mathbb{R}^{1440}$  
\REPEAT
\STATE Assign each feature vector to clusters based on the closest Euclidean norm.
\STATE Update the position of the centroids based on their mean distances to assigned points.
\UNTIL Clustering converged
\ENDFOR
\STATE Obtain the optimal $K$ denoted as $K_c$ using Elbow method.
\STATE Initialize: $K_c$ clustering centroids $\mu_1, \mu_2 \in \mathbb{R}^{1440}$
\REPEAT
\STATE Assign each feature vector to clusters.
\STATE Update the positions of the centroids.
\UNTIL Clustering converged

\end{algorithmic}

\caption{k-means sensors clustering}
\label{alg:kmean}
\end{algorithm}
\end{frame}

\section{Results}\label{sec:results}

In this section, we evaluate our proposed model using traffic flow data obtained from the Caltrans Performance Management System (PeMS) \citep{PeMS}.  \textcolor{black}{We first evaluate the imputation performance for a single sample sensor followed by large-scale districtwide sensors. Then, we show the efficiency of our proposed model by comparing it with other benchmark baseline models, namely support vector regression (SVR), history average (HA), denoising stacked autoencoder (DSAE), and GAN based parallel data model \citep{chen2019traffic}.}

\subsection{Data Description}\label{sec:datadescription}
The Caltrans PeMS dataset used in this study, is one the most popular open source dataset for transportation research, consisting of more than 15,000 vehicle detector stations (VDSs) or sensors covering over the entire state of California. In this study, we used the 5-minute traffic flow data provided by the PeMS data warehouse for the year 2013 from District 5: Central Coast. There were 147 VDSs in this district, each of which had 363 days’ worth of traffic flow data vectors, while no data was present for the remaining two days of the year. Hence, each individual VDS had 104,544 ($363\times24\times12$) traffic records. We divided the weekday data vectors and non-weekday data vectors following the work proposed by \cite{duan2016efficient} and \cite{chen2019traffic}. This resulted in 245 days labeled as weekdays and 118 days labeled as non-weekdays for each individual VDS. It should be noted that our dataset is exactly the same dataset used in the \cite{duan2016efficient} and \cite{chen2019traffic} study for DSAE model and GAN based parallel data model respectively. This enabled us to directly compare the performance of our model with these benchmark models.

\subsection{Evaluation Criteria}\label{sec:criteria}

In order to evaluate the performance of our TSDIGAN model, we utilized three criteria: mean absolute error (MAE), root mean square error (RMSE), and mean relative error (MRE), given by the following equations:

\begin{equation}\label{eq:mae}
MAE = \frac{{\sum\nolimits_{d = 1}^{D_{test}} {\sum\nolimits_{t = 1}^T {{I^s_{d,t}}|{y_{d,t}} - {{\hat y}_{d,t}}|} } }}{{\sum\nolimits_{d = 1}^{D_{test}} {\sum\nolimits_{t = 1}^T {{I^s_{d,t}}} } }}
\end{equation}
\\
\begin{equation}\label{eq:rmse}
RMSE = \sqrt {\frac{{\sum\nolimits_{d = 1}^{D_{test}} {\sum\nolimits_{t = 1}^T {{I^s_{d,t}}{{({y_{d,t}} - {{\hat y}_{d,t}})}^2}} } }}{{\sum\nolimits_{d = 1}^{D_{test}} {\sum\nolimits_{t = 1}^T {{I^s_{d,t}}} } }}} 
\end{equation}
\\
\begin{equation}\label{eq:mre}
MRE = \frac{{\sum\nolimits_{d = 1}^{D_{test}} {\sum\nolimits_{t = 1}^T {{I^s_{d,t}}\frac{{|{y_{d,t}} - {{\hat y}_{d,t}}|}}{{{y_{d,t}}}}} } }}{{\sum\nolimits_{d = 1}^{D_{test}} {\sum\nolimits_{t = 1}^T {{I^s_{d,t}}} } }}
\end{equation}

where, $y_{d,t}$ is the observed traffic flow data (groundtruth), while $\hat y_{d,t}$ is the imputed traffic flow data obtained using the proposed model. $D_{test}$ is the total number of daily traffic flow vectors used for testing, $T$ is the dimension of each traffic flow vector (equal to 288), and $I^s_{d,t}$ is the corrupted mask mentioned in Section \ref{sec:notation}.

To fairly evaluate our model’s performance throughout the next steps of our study, we randomly corrupted the observed data with various random missing rates (MR), and distributed the missing data points equally for each test sample. The random missing rate (MR) can be defined as:

\begin{equation}
MR = \frac{{\sum\nolimits_{d = 1}^{{D_{test}}} {\sum\nolimits_{t = 1}^T {{I^s_{d,t}}} } }}{{{D_{test}} T}} \times 100\%
\end{equation}

For the convenience of evaluation and comparison in the following sections, we used a single default ratio of 4:1 to split the training and test samples \textcolor{black}{ for each individual sensor. Therefore, 245 weekdays were split into 196 days for training and 49 days for testing, while the remaining 118 non-weekday patterns were split into 94 days for training and 24 days for testing.} And for each MR, we repeated the experiments 25 times and took the average to ensure unbiased and reliable results. Next, we describe our model performance on a single sample sensor.

\subsection{Single VDS Performance}\label{sec:single_performance}
 
\begin{figure*}[!htb]
    \centering
    \includegraphics[width=0.95\textwidth]{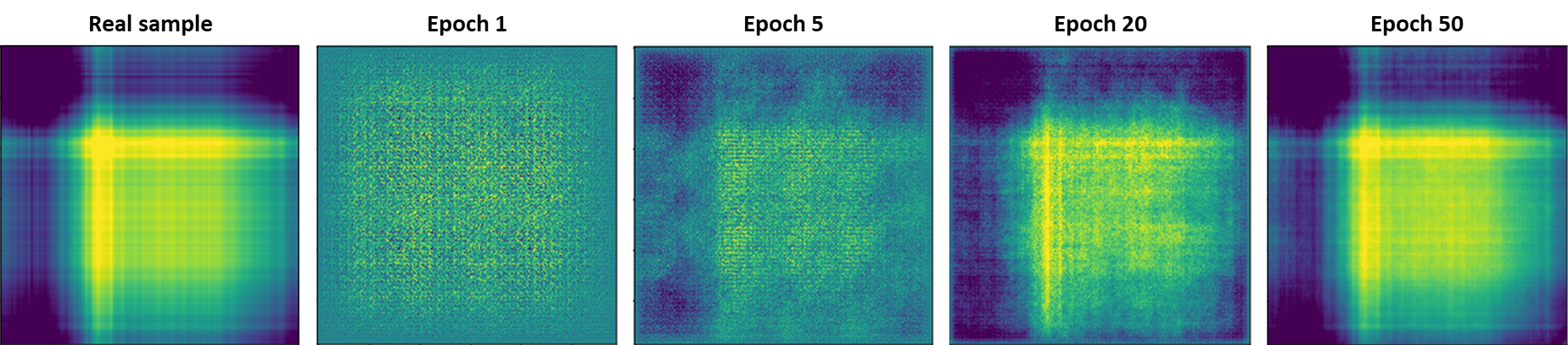}
    \caption{Sample results of GASF encoded images training using the proposed GAN generator model.}
    \label{fig:training}
\end{figure*}

As mentioned in Section \ref{sec:methodology}, we trained our proposed model using fully observed training samples. Figure \ref{fig:training} shows the training process of the proposed model across different epochs along with the real observed data. The generator tends to learn to create realistic-looking GASF matrix images step by step. After 50 epochs, the generator was able to produce synthetic GASF images that are similar to the real samples, as shown in the left-most sub-figure of Figure \ref{fig:training}. We then used the imputation module described in Section \ref{sec:TSDIGAN_imputation} to impute our corrupted test samples from the optimal latent space. Here, we used the VDS 500010102 as our target VDS for demonstration, which is the same used in \cite{duan2016efficient}. As mentioned in Section \ref{sec:datadescription}, we trained separate models for weekdays and non-weekdays. Therefore, we had 196/49 traffic flow data vectors for training/testing for weekdays, and 94/24 traffic flow data vectors for training/testing for non-weekdays. We then stacked our imputed result vectors $\hat y_{d,t}$ from both weekdays and non-weekdays together to evaluate the overall accuracy using the criteria equations mentioned in Section \ref{sec:criteria}. This setup was used by default for all later experiments.  

\begin{figure}[!htb]
    \centering
    \includegraphics[width=0.6\textwidth]{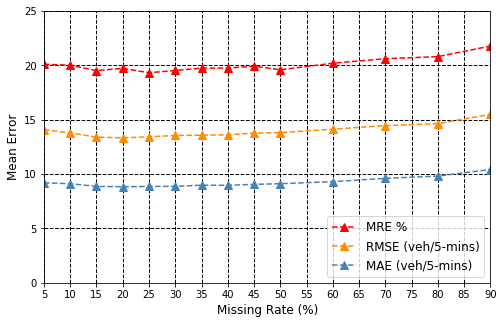}
    \caption{Imputation performance for a single sample sensors for different missing rates.}
    \label{fig:single_fig}
\end{figure}

We conducted experiments to test the efficiency and robustness of our proposed model for MRs ranging from 5\% to 90\%. The results obtained for the sample sensor is shown in Figure  \ref{fig:single_fig}. $MAE$ was found to vary between 8.8 to 10.4 vehicles per 5 minutes (veh/5-mins), $RMSE$ from 13.3 to 15.4 veh/5-mins, while $MRE$ ranged from 19.3\% to 21.7\%. As it can be seen from the figure, while the error trace increases with increase in $MR$, however, our proposed approach was able to perform reasonably well even in very high $MR$ ($\ge$ 50\%).  Even under  $MR$ as high as 80\%, our proposed model was still able to obtain decent imputation results with an $MAE$ of less than 10.0 veh/5-mins and $MRE$ of less than 21.0\%. This shows that the proposed model is robust to high missing data percentages too.

\begin{figure*}[!htb]
    \hspace{+10 pt}
    \includegraphics[width=0.9\textwidth]{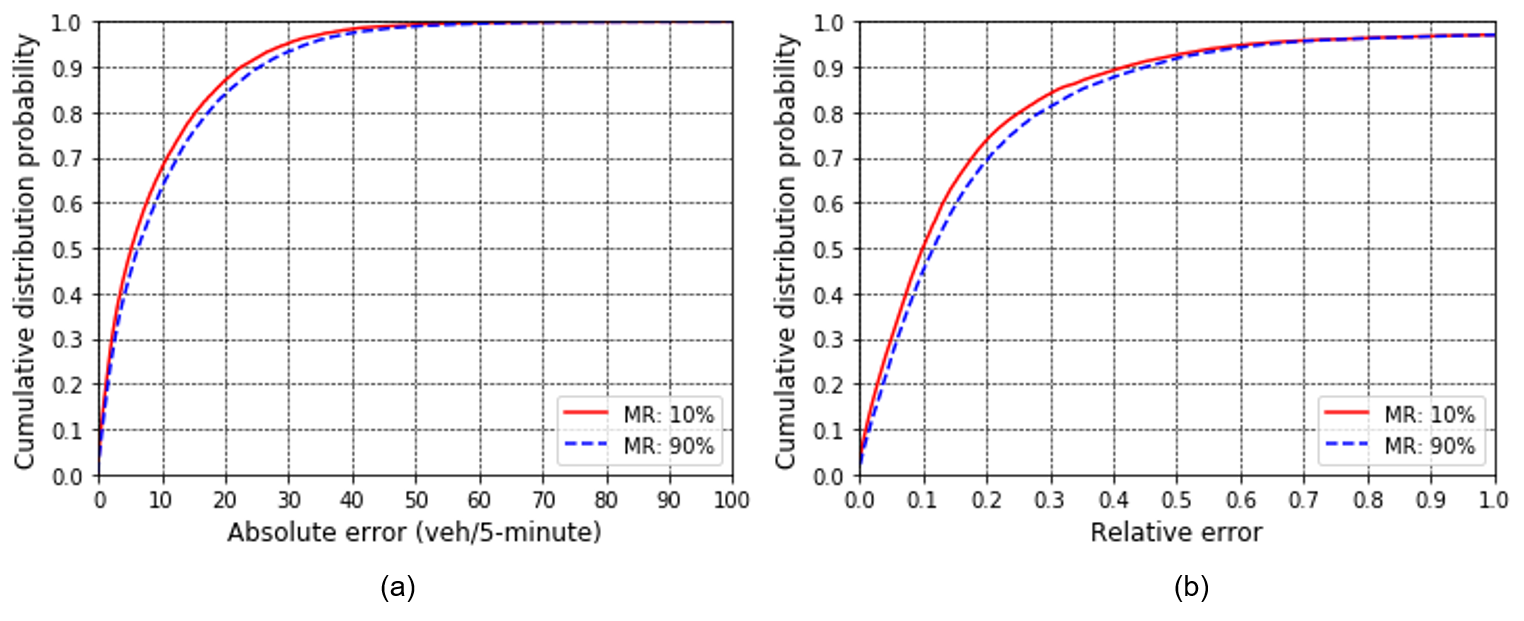}
    \caption{Error distributions for 10\% and 90\% marginal missing rates: (a) absolute error and (b) relative error.}
    \label{fig:dist}
\end{figure*}

\begin{figure*}[!htb]
    \centering
    \begin{subfigure}[b]{0.65\textwidth}
        \hspace{-5 pt}
        \includegraphics[height=1.65in]{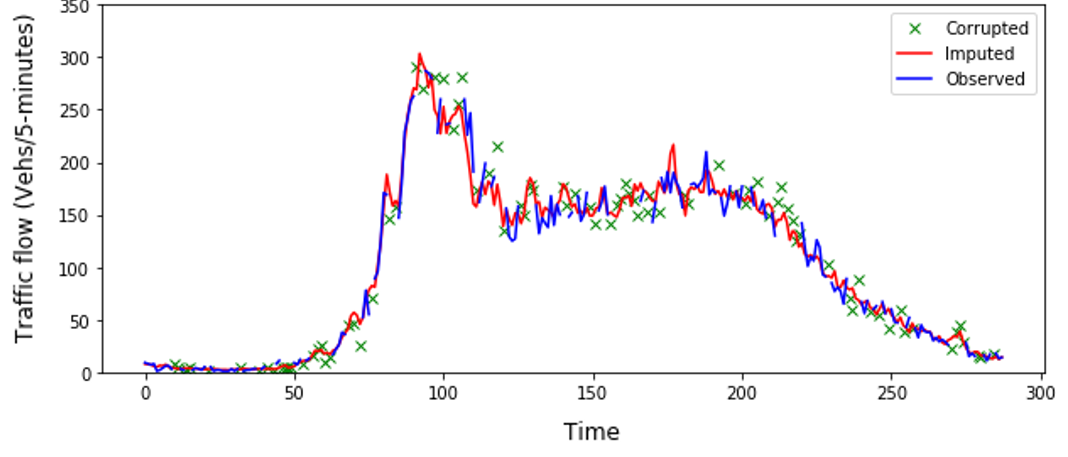}
        \caption{}\label{fig:single_plotw}
    \end{subfigure}
    
    \begin{subfigure}[b]{0.65\textwidth}
        \hspace{-5 pt}
        \includegraphics[height=1.65in]{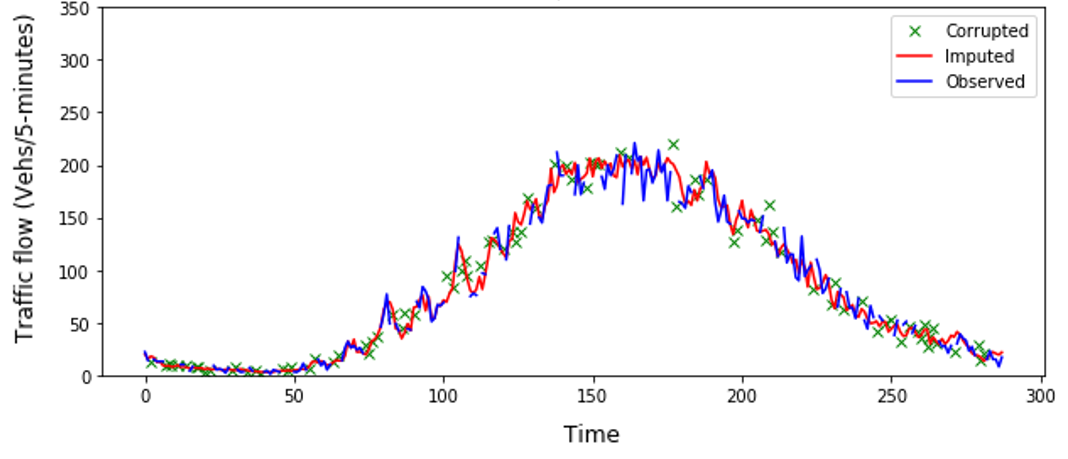}
        \caption{}\label{fig:single_plotn}
    \end{subfigure}   

    \caption{Sample synthetic traffic flow data plot: (a) Weekday and (b) Non-weekday.}\label{fig:single_plot}
\end{figure*}

To extend this further, Figure  \ref{fig:dist} shows the absolute error ($AE$) distribution and the relative error ($RE$) distributions for 10\% and 90\% $MR$s. This can help to understand the $AE$ and $RE$ variation range within two extreme $MR$s (10\% and 90\%). It can be observed that about 60\% of $AE$ was less than 8 veh/5-mins at 10\% $MR$, while it is less than 9 veh/5-mins at 90\% $MR$. Similarly, there is about 60\% of $RE$ less than 13.5\% at 10\%  $MR$, while it is less than 15.5\% at 90\% $MR$. By taking advantage of the generative model and temporal dependency correlation GASF matrix, our proposed model can produce robust and reliable imputation results even for large variation of $MR$s. A sample weekday and non-weekday imputation results is shown in Figure \ref{fig:single_plot}a and \ref{fig:single_plot}b, respectively. The figure shows the imputed data obtained using the proposed model along with the corresponding actual ``observed" data and corrupted data too. It can be seen that the sample synthetic traffic flow data produced by our proposed model perform reliably well in successfully overlaying the observed part, with its ``closest” estimation of the corrupted data points.

\begin{figure*}[!htb]
    \centering
    \includegraphics[width=0.95\textwidth]{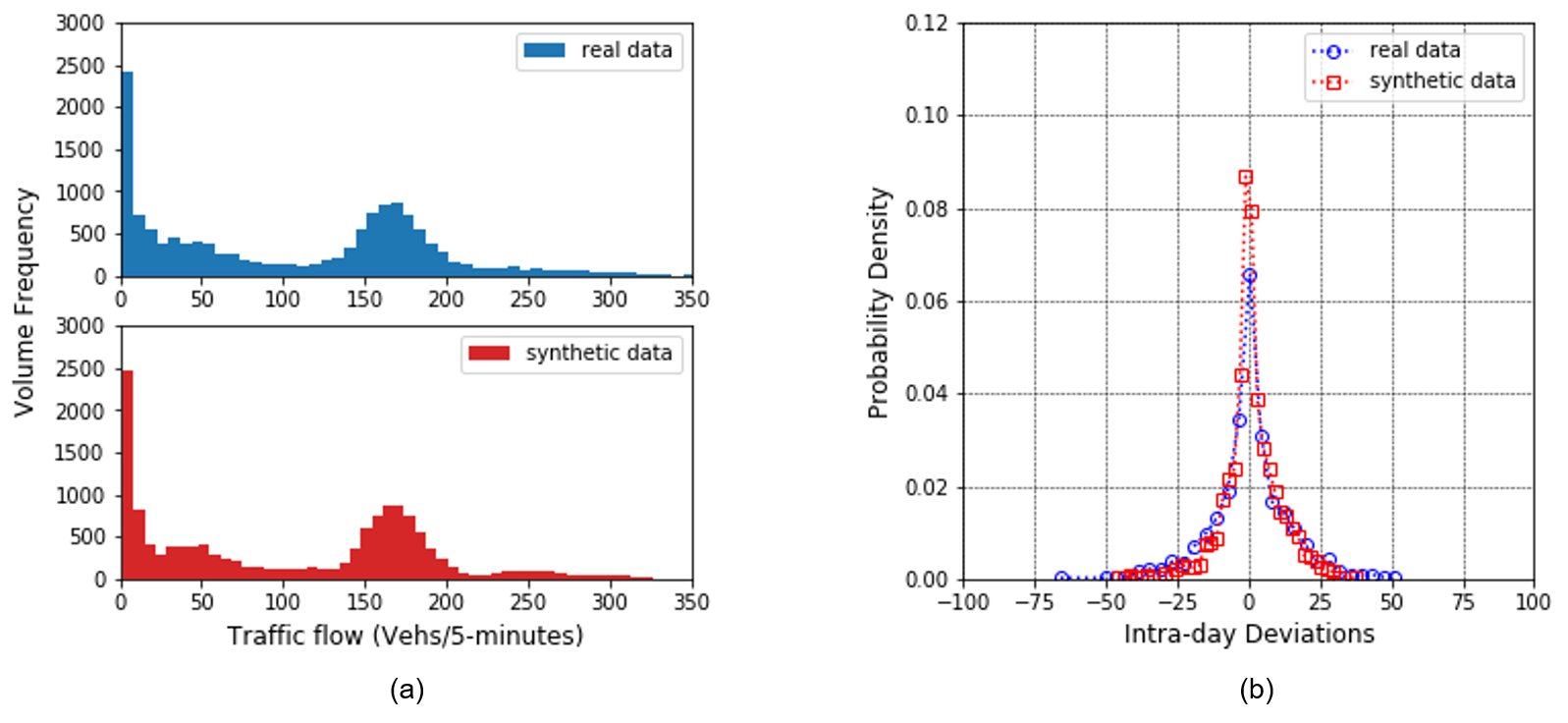}
    \caption{ (a) Traffic flow histograms and (b) deviation distribution between the real test data and synthetic data for 20\% MR .}
    \label{fig:distribution_plot}
\end{figure*}

\textcolor{black}{To illustrate the efficiency of our proposed model in learning the traffic data distribution, we plot the traffic flow histograms generated using the real data and the synthetic data obtained from the model for 20\% $MR$, as shown in Figure \ref{fig:distribution_plot}a. Further, Figure \ref{fig:distribution_plot}b shows the empirical distributions of the deviation time series of the real data and the synthetically generated data, similar to \cite{chen2012retrieval, li2013efficient}. The deviations are calculated as difference between simple average intra-day trend from the original and imputed data. This helps to check if the imputed data preserve the important statistical features of the original dataset. As it can be seen from  Figure \ref{fig:distribution_plot}(a) and (b), the synthetic data distributions closely follow the original data distributions, thereby verifying that the proposed imputation technique has been able to retain the original data features successfully. In the next section, we discuss the details of our proposed model performance in a large-scale implementation over the entire District 5 of California instead of a single sensor imputation.}

\subsection{Imputation Performance on a Large-Scale Network}\label{sec:state_implementation}

\begin{figure}[!htb]
    \centering
    \includegraphics[width=0.45\textwidth]{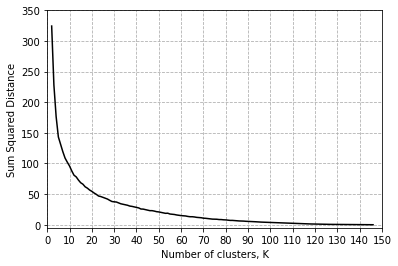}
    \caption{Elbow plot to determine optimal number of clusters for VDSs}
    \label{fig:elbow}
\end{figure}

In this subsection, we evaluate our proposed model using the entire 2013 data obtained from all VDSs of District 5 of California. As mentioned in Section \ref{sec:TSDIGAN_state}, we first divided the 147 VDSs into different homogeneous groups based on their daily traffic flow patterns using the k-means and elbow method. As shown in Figure \ref{fig:elbow}, we draw the sum of squared distances versus the possible number of clusters, and used the elbow method to determine the optimal number of clusters ($K_c$) \citep{kodinariya2013review}. The optimal number of clusters were found to be 25 using the elbow point. However, the selection of the optimal groups can also be chosen based on the agency specific requirements.

\begin{figure*}[!htb]
    \centering
    \begin{subfigure}[b]{0.49\textwidth}
        \centering
        \includegraphics[height=2.30in]{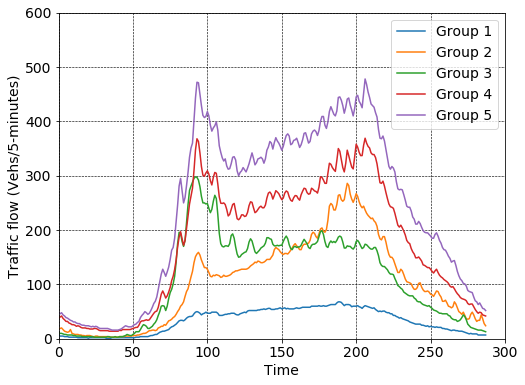}
        \caption{\centering Weekday}\label{fig:single_plotwg}
    \end{subfigure}
    ~
    \begin{subfigure}[b]{0.49\textwidth}
        \centering
        \includegraphics[height=2.30in]{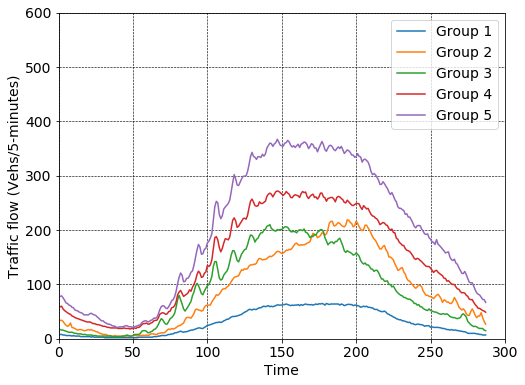}
        \caption{\centering Nonweekday}\label{fig:single_plotng}
    \end{subfigure}   
    \caption{Median traffic flow patterns for (a) weekdays and (b) non-weekdays for 5 sample VDS clusters}\label{fig:group_pattern}
\end{figure*}

To demonstrate the different daily patterns observed in the generated clusters, we plot the median daily traffic flow data of 5 sample clusters in Figure \ref{fig:group_pattern}. The average daily traffic (ADT) for these 5 sample groups varied between 9,000 to 64,000 vehicles. It can be seen that there were both morning and evening peaks for groups 4 and 5, while groups 2 and 3 have either a morning peak or an evening peak. It can also be seen that group 1 had the lowest daily traffic flow. Therefore, the clusters generated having distinct traffic flow patterns help the model to learn those unique patterns and perform more efficiently in large-scale network. 

\begin{figure*}[!htb]
    \centering
    \includegraphics[width=0.95\textwidth]{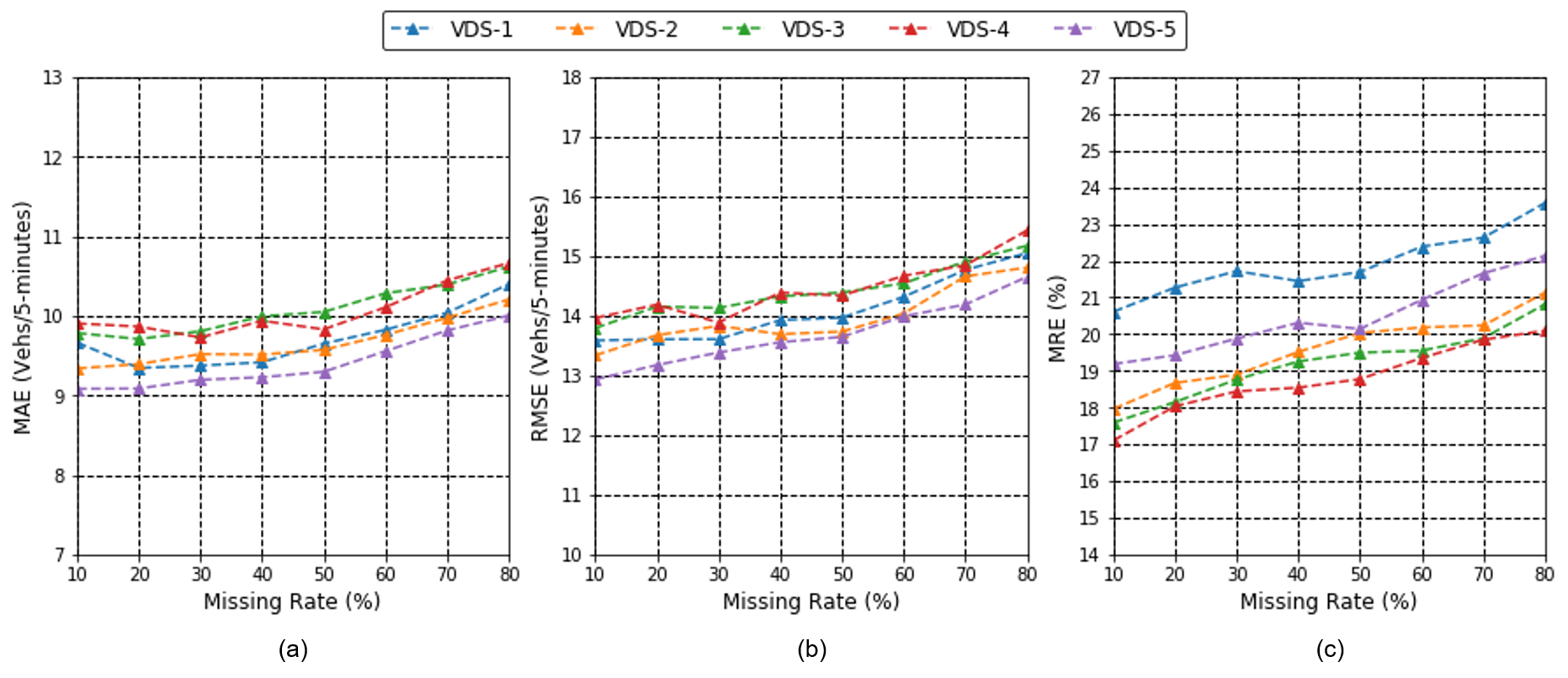}
    \caption{Imputation performance in terms of  (a) MAE, (b) RMSE, and (c) MRE for all VDS in Cluster ID 2}
    \label{fig:group_fig}
\end{figure*}

In addition, to ensure fair and balanced results for each individual VDS, we selected 20\% of the samples equally from each VDS for testing and used the remaining 80\% for training. Figure  \ref{fig:group_fig} shows the performance of our proposed model for all VDS in a sample group 2 for MR ranging between 10\% to 80\%. It can be seen that $MAE$ for all VDSs ranging from 9.1 to 10.6 veh/5-mins,  $RMSE$ ranging from 12.9 to 15.3 veh/5-mins, and $MRE$ ranging from 17.1\% to 23.6\%. This shows that the proposed model scales efficiently for larger number of VDSs, in addition to the single VDS model described in Section \ref{sec:single_performance}.  

Table \ref{tab:group_tab} summarizes the overall performance of our proposed TSDIGAN model for the 5 sample clusters, whose daily pattern is shown in Figure \ref{fig:group_pattern}. It can be seen that the model performed reasonably well even in high $MR$ of 80\%, showing the efficacy  of the model. Further, although the $MAE$ and $RMSE$ increased for clusters with higher $ADT$, the $MRE$ showed decreasing trend with increase in $ADT$, thereby showing it's performance is robust for different ranges of $ADT$ too.

\begin{table*}[!htb]
  \centering
  \caption{Performance summary for VDSs of 5 sample clusters at 20\%, 50\%, and 80\% MR}
    \begin{tabular}{llccrcccrcccrc}
    \toprule
    \multicolumn{2}{l}{Criteria} &       & \multicolumn{3}{c}{MAE (veh/5-min)} &       & \multicolumn{3}{c}{RMSE (veh/5-min)} &       & \multicolumn{3}{c}{MRE} \\
\cmidrule{1-2}\cmidrule{4-6}\cmidrule{8-10}\cmidrule{12-14}    Group & ADT   &       & 20\%  & \multicolumn{1}{c}{50\%} & 80\%  &       & 20\%  & \multicolumn{1}{c}{50\%} & 80\%  &       & 20\%  & \multicolumn{1}{c}{50\%} & 80\% \\
    \midrule
    1     & 9000  &       & 5.0   & 5.2   & 5.6   &       & 7.2   & 7.5   & 8.1   &       & 0.318 & 0.327 & 0.350 \\
    2     & 29000 &       & 9.5   & 9.7   & 10.4  &       & 13.8  & 14.0  & 15.0  &       & 0.191 & 0.200 & 0.216 \\
    3     & 30000 &       & 9.8   & 10.0  & 10.6  &       & 14.5  & 14.9  & 15.5  &       & 0.200 & 0.210 & 0.221 \\
    4     & 47000 &       & 13.9  & 14.1  & 15.2  &       & 19.8  & 20.4  & 21.7  &       & 0.119 & 0.124 & 0.133 \\
    5     & 64000 &       & 14.6  & 14.9  & 16.4  &       & 20.7  & 21.4  & 22.9  &       & 0.103 & 0.108 & 0.116 \\
    \bottomrule
    \end{tabular}%
  \label{tab:group_tab}%
\end{table*}%

\begin{figure*}[!htb]
    \centering
    \includegraphics[width=0.65\textwidth]{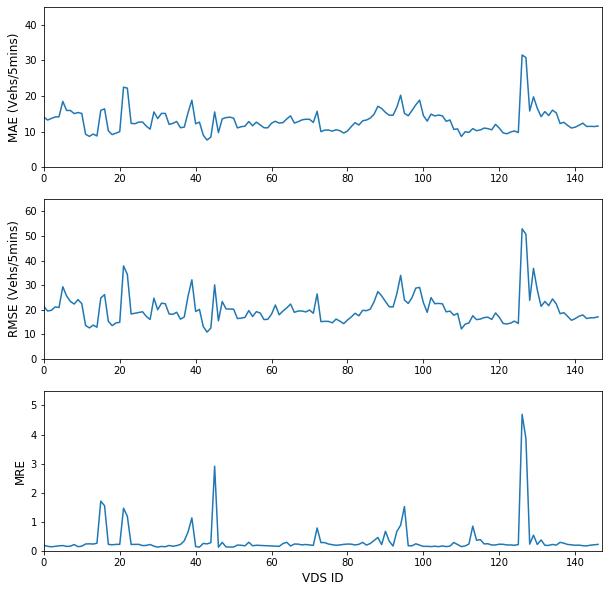}
    \caption{Imputation performance accuracies ($MAE$, $RMSE$, and $MRE$) at 30\% MR for all VDSs}
    \label{fig:group_performance}
\end{figure*}

\textcolor{black}{Figure \ref{fig:group_performance} presents the  imputation performance accuracies in terms of $MAE$, $RMSE$, and $MRE$ for all 147 VDSs of the entire District 5 of California at 30\% MR. For all VDSs,  $MAE$ was found to vary between 7.61 to 31.55 veh/5-mins with the median and mean values of 12.64 and 13.16 veh/5-mins respectively. Similarly, $RMSE$ varied between 10.91 to 52.9 veh/5-mins with the median and mean of 18.98 and 20.22 veh/5-mins respectively. Therefore, the proposed model performed reasonably well across large-scale sensor networks in terms of $MAE$ and $RMSE$.  However,  $MRE$ was found to vary between 12.4\% to 469\% with the median and mean values of 20.7\% and 35.5\%.} 
\textcolor{black}{Therefore, the $MRE$ performance for the proposed model was found to be highly skewed, with exceptionally high $MRE$ for VDS ID 126 and 127. On further investigation, it was observed that the input of raw training data of these sensors had more inherent noise and zero volume report 
which made learning of the sufficient representative ``pattern" from such training data highly unstable. However, the median $MRE$ across all sensors were found to be only 20.7\%, suggesting reasonable overall performance across the sensors. These results indicates that the efficiency of our proposed TSDIGAN framework across districtwide sensor networks, thereby showing it's feasibility for large-scale practical implementations.}

\textcolor{black}{Additionally, efficiency of real-world implementation of the proposed model depends on the training and testing cost and time requirements. This is particularly important since training deep learning models is time-consuming and GANs in particular are well-known to suffer from vanishing gradients, mode collapse, and failure to convergence. Our proposed model was trained and tested using a single NVIDIA GTX 1080Ti GPU along with Intel(R) i7-8700 CPU, 32 GB RAM, and Windows 10 (64 bits) platform. All the frameworks used in this study were built using PyTorch 1.1 \citep{paszke2017automatic}. Our proposed model took approximately 8 minutes training time for a single VDS and 14 hours for the entire districtwide 147 VDSs using one year of historical traffic flow data for 50 epochs. The average training time for each group of VDS obtained using clustering method is found to be around 30 minutes. To find the optimal latent space using the imputation module during the test phase, the time taken varies depending on the iterations. In our experiments, we performed 200 iterations which was found to take approximately 2 seconds for generating daily traffic data generation for a single VDS. Therefore, it takes approximately 5 minutes for districtwide daily traffic data imputation across 147 VDS. This shows that the proposed model can be successfully applied to large-scale real-world implementation scenarios with the desired regular offline model retrain/update.  Next, we present the results on comparison of our proposed model with other benchmark data imputation models.}

\subsection{Model Comparison}\label{sec:compare_performance}

\begin{figure*}[!htb]
    \centering
    \includegraphics[width=0.65\textwidth]{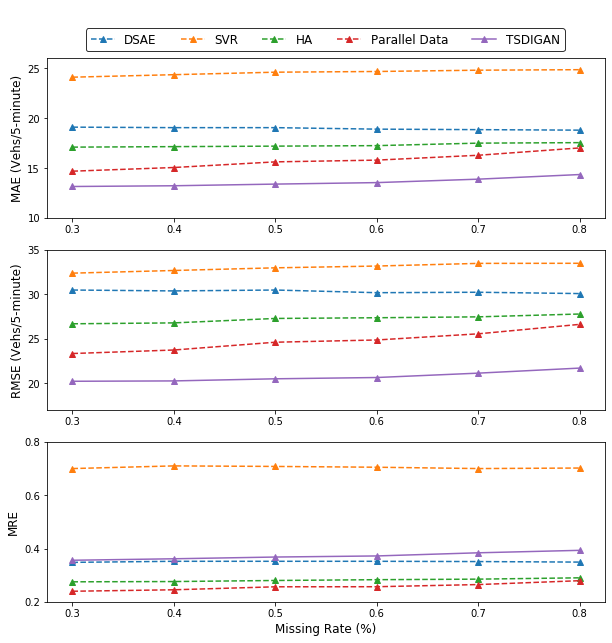}
    \caption{Comparison of imputation performance accuracies in terms of (a) $MAE$, (b) $RMSE$, and (c) $MRE$ with respect to other benchmark imputation models}
    \label{fig:compare_fig}
\end{figure*}

In this section, we compare the performance of our proposed TSDIGAN model with other benchmark traffic data imputation models to find out the efficiency of our proposed model. \textcolor{black}{The benchmark models used in this study for comparison are support vector regression (SVR), history average (HA), denoising stacked autoencoder (DSAE), and GAN based parallel data model \citep{chen2019traffic}. It should be noted that the benchmark dataset used in our study was also used by \cite{duan2016efficient} for DSAE model and \cite{chen2019traffic} for GAN based parallel data model. This enabled us to directly compare the performance of our model with these benchmark models. The detailed default settings for model training and evaluation results for baseline models can be found in \cite{chen2019traffic}. Figure \ref{fig:compare_fig} shows the average imputation performance accuracies across all VDS in terms of $MAE$, $RMSE$, and $MRE$ for all comparison models. It can be seen that our proposed model outperformed all other benchmark models in terms of $MAE$ and $RMSE$. An overall improvement of 13.7 \% and 16.3 \% was observed by TSDIGAN model compared to the next best performing parallel data model. However, in terms of $MRE$, the proposed TSDIGAN model performed poorly to other benchmark models, except SVR. This can be contributed due to a few sensors for which $MRE$ was found to be significantly higher, as shown in Figure \ref{fig:group_performance} and described in Section \ref{sec:state_implementation}. Further, while the benchmark models were trained individually for each sensor separately, we trained our models for each cluster or group of sensors  which can be attributed to be one of the reason why our model performs poorly for sensors with significant noise or zero volume report compared to other sensors. It can be pointed out that while the mean $MRE$ for TSDIGAN across all sensors for 30\% MR was found to be 35.5\%, the median $MRE$ was only 20.7\%. Also, the mean $MRE$ for 95\% of sensors was found to vary between 24.0\% - 26.1\% in comparison to 35.5\% - 39.4\% variation, when all sensors are considered. This implies that the mean value shown in Figure \ref{fig:compare_fig} was significantly affected by performance on few outlying sensors, which led to its poor performance compared to other benchmark models. In future, more efficient clustering techniques can be used either to remove such sensors from performance analysis or separate models can be trained for such sensors, depending on user specific requirements. Overall, the proposed TSDIGAN model outperformed all benchmark models in terms of $MAE$ and $RMSE$, while performing reasonably well in terms of $MRE$ too for majority of sensors.}

\section{Conclusion}\label{sec:conclusion}

In this study, we propose a traffic sensor data imputation framework based on generative adversarial networks (TSDIGAN) that treats the missing data problem as a data generation problem. Our study demonstrates that the generative model based method can perform accurately and robustly to impute missing traffic data under widely varying missing rates. Our proposed model first embeds traffic time-series data into GASF matrix images preserving the temporal correlations. This enables training of a deep convolutional generative adversarial network that can generate realistic-looking synthetic data for missing data imputation. We have also shown our proposed model’s training process step by step, demonstrating how our model learns to generate its high-quality synthetic data. We have evaluated the performance of the proposed model using benchmark data from PeMS \citep{PeMS} and further investigated it's capability for large-scale applications. We compared our proposed model performance with other benchmark models, including  \textcolor{black}{support vector regression (SVR), history average (HA), denoising stacked autoencoder (DSAE), and GAN-based parallel data model. Our results show that the proposed model can outperform the benchmark models in terms of $MAE$ and $RMSE$, while achieving comparable accuracies in terms of $MRE$ for majority of the sensors. Further, our proposed framework groups the sensors into clusters based on the similarity of their daily traffic patterns to learn the generative model which can be applied to the entire cluster. This can help to train fewer cluster-specific models instead of maintaining each sensor specific model, thereby handling the entire training, testing, and real-world application procedure more efficiently.}

Our proposed framework can easily and cheaply generate a variety of realistic synthetic traffic data, which makes it a good choice when it is inconvenient or impossible to get sufficient real traffic data. In addition, the characteristics of our proposed framework offer the possibility of extended ITS applications like data analysis enhancement, anomaly detection, etc. In future, this can be integrated with external features such as weather, special events, and other factors that can impact traffic flow patterns to enable our model to provide more adaptive and accurate imputation performance to appropriately reflect different conditions. \textcolor{black}{Further, in this study, we used k-means clustering to group the sensors based on their daily traffic patterns and develop models for each cluster. In future, this study can be extended to evaluate other efficient clustering techniques such as hierarchical clustering, density based clustering and even determining optimal variation of temporal and spatial traffic data characteristics which can be grouped and worked upon as a single cluster.} Also, this can be extended to evaluate the suitability and effectiveness of such generative model based deep learning frameworks for traffic speed generation, prediction, and similar other ITS applications.

\section*{Acknowledgements}
Our research results are based upon work supported by the Iowa DOT Office of Traffic Operations Support Grant. Any opinions, findings, and conclusions or recommendations expressed in this material is of the author(s) and do not necessarily reflect the views of the Iowa DOT Office of Traffic Operations.

\bibliography{main.bib}

\end{document}